\documentclass[preprint,12pt,eqsecnum,aaspp4]{aastex}
\usepackage{natbib,graphics}

\newcommand{\be}{\begin{equation}}
\newcommand{\ee}{\end{equation}}

\begin{document}
\slugcomment{Submitted to ApJ}

\title{Observational Constraints on the Self Interacting Dark Matter Scenario and the Growth of Supermassive Black Holes}
\author{Joseph F. Hennawi}
\email{jhennawi@astro.princeton.edu}
\and
\author{Jeremiah P. Ostriker}
\email{jpo@astro.princeton.edu}
\affil{Princeton University Observatory, Princeton, NJ 08544}

\begin{abstract}

	We consider the astrophysical consequences of the self interacting dark matter (SIDM) scenario for a general velocity dependent cross section per unit mass which varies as some power of velocity: $\sigma_{DM}=\sigma_0\left(v/v_0\right)^{-a}$.  Accretion of SIDM onto seed black holes can produce supermassive black holes that are too large for certain combinations of $\sigma_0 v_0^a$ and $a$, a fact which is used to obtain a new constraint on the dark matter interaction.  Constraints due to other astrophysical considerations are presented and previous constraints for a constant cross section are generalized. The black hole constraint is extremely sensitive to the cusp slope $\alpha$, of the inner density profile $\rho \sim r^{-\alpha}$ of dark halos.  For the most probable value of $\alpha=1.3$, we find that there exists a tiny region in the parameter space for SIDM properties, with $a \approx 0.5$ and $\left(\sigma_0/1 \ \mathrm{cm}^2 \ \mathrm{g}^{-1}\right)\left(v_0/100 \ \mathrm{km} \ \mathrm{s}^{-1}\right)^a \approx 0.5$, such that all constraints are satisfied.  However, the adiabatic compression of the dark halo by baryons as they cool and contract in normal galaxies yields a steeper cusp, $\rho \sim r^{-\alpha^{\prime}}$.  We find that in both the highly collisional and collisionless limits, invariance arguments require $\alpha^{\prime}=\frac{\alpha+3\xi-\alpha\xi}{4-\alpha}$, where $\alpha$ and $\alpha^{\prime}$ are the inner profile slope of the dark halo before and after compression, respectively.  This gives the tighter constraint  $\left(\sigma_0/1 \ \mathrm{cm}^2 \ \mathrm{g}^{-1}\right)\left(v_0/100 \ \mathrm{km} \ \mathrm{s}^{-1}\right)^a \lesssim 0.02$, which would exclude SIDM as a possible solution to the purported problems with CDM on subgalactic scales in the absence of other dynamical processes.  Nevertheless, SIDM with parameters consistent with this stronger constraint, can explain the ubiquity of supermassive black holes in the centers of galaxies.  A ``best fit'' model is presented with $a=0$ and $\left(\sigma_0/1 \ \mathrm{cm}^2 \ \mathrm{g}^{-1}\right)= 0.02$,  which reproduces the supermassive black hole masses and their observed correlations with the velocity dispersion of the host bulges.  Specifically, the approximately fourth power dependence of black hole mass on galactic velocity dispersion is a direct consequence of the power spectrum of primeval perturbations having an index of $n \approx -2$ and the value of $a$.  Although the dark matter collision rates for this model are too small to directly remedy problems with CDM, mergers between dark halos harboring supermassive black holes at high redshift could ameliorate the cuspy halo problem.  This scenario also explains the lack of comparable supermassive black holes in bulgeless galaxies like M33.

\end{abstract}


\section{Introduction}

	Self interacting dark matter (SIDM) has been proposed by Spergel \& Steinhardt (2000) to remedy purported problems with the cold dark matter (CDM) family of cosomological models on subgalactic scales .  The apparent conflicts between numerical simulations of CDM and observations from dwarf galaxies to clusters seem to indicate that CDM halos are too centrally concentrated (see Wandelt et al. 2000 and references therein).  Of these discrepancies, two problems are particularly significant.  First, the inner density profiles of CDM halos diverge as $\rho \propto r^{-\alpha}$ with $\alpha \approx 1.3\pm 0.2$ (Navarro, Frenk, \& White 1996, 1997; Fukushige \& Makino 1997; Moore et al. 1999; Subramanian, Cen, \& Ostriker 2000; Jing \& Suto 2000; Ghigna et al. 2000; Klypin et al. 2001; Fukushige \& Makino 2001a, 2001b), possibly conflicting with profiles deduced from rotation curve observations of dark matter dominated dwarf and low surface brightness galaxies (Flores \& Primack 1994; Dalcanton \& Bernstein 1997; De Blok \& McGaugh 1997; Swaters, Madore \& Trewhella 2000; though see van den Bosch et al. 2000 and van den Bosch \& Swaters 2001).  Second, numerical simulations of the CDM scenario indicate an excess number of small scale structures when compared with the number of dwarf galaxies in the Local Group or satellites in galactic halos (Moore et al. 1999; Klypin et al. 1999).  Endowing dark matter with a finite cross section for elastic scattering allows heat transfer to occur over a Hubble time, resulting in less concentrated structures, flattened density profiles, and fewer substructure satellites (Burkert 2000; Yoshida et al. 2000b; Dav\'{e} et al. 2000; Wandelt et al. 2000). 

	If $\sigma_{DM}\equiv \sigma_p/m_p$ is the cross section per unit mass of a dark matter particle, then the mean free path is $\lambda = 1/\rho\sigma_{DM}$, where $\rho$ is the density of dark matter.  An optical depth can be defined $\tau \equiv r/\lambda$, which distinguishes different physical regimes.  In regions which are optically thick $\tau \gg 1$, the dark matter behaves as a fluid; whereas, for regions which are optically thin $\tau \lesssim 1$, the dynamics closely resembles two body relaxation in globular clusters.  The range of cross sections consistent with experimental constraints, yet giving the required rates of evolution, imply that dark halos will probably be optically thin $\tau\left(r_s\right)\lesssim 1$ at their characteristic scale radius $r_s$ (Wandelt et al. 2000; Dav\'{e} et al. 2000), but of course can be optically thick in the inner regions $r\ll r_s$ where the densities are higher (provided $\alpha>1$).  

	Numerical investigations of SIDM halo evolution have been carried out by several groups.  Yoshida et al. (2000a) and Moore et al. (2000) simulated SIDM halos in the optically thick or fluid limit, which, as mentioned above, is probably not the relevant scenario for SIDM.  Burkert (2000) and Kochanek \& White (2000) simulated isolated halos employing a range of cross sections nearly consistent with the optically thin requirement.  Both groups found that halos develop shallow cores for a modest period of time before the onset of core collapse, but disagree on the core collapse timescale.  Dav\'{e} et al. (2000) and Yoshida et al. (2000b) simulated optically thin halos in cosmological settings, which include infall and merging. They observed evolution towards reduced central densities and shallower inner profiles, effects increasing with increasing cross section; and neither saw any evidence for core collapse. Their results are in broad agreement with each other and span the range from dwarf galaxies to clusters.  Alternative to N-body techniques, Hannestead (2000) and Firmani, D'Onghia, \& Chincarini (2001) have carried out integrations of the collisional Boltzmann equation. Firmani et al.'s simulation differs from previous investigations in that they considered a velocity dependent cross section $\sigma \propto v^{-1}$.  Also Balberg, Shapiro \& Inagaki (2001) examined SIDM halo evolution via a time-dependent gravothermal numerical calculation. All three find the development of shallower central slopes and less concentrated cores than CDM would produce, consistent with the quoted N-body results, although Balberg et al. (2001) obtained a collapse timescale roughly an order of magnitude larger than that seen by Burkert (2000) and Kochanek \& White (2000).  In sum, the most relevant numerical studies performed to date indicate that, for appropriate dark matter scattering cross sections, collisions can effectively reduce dark matter central densities for normal galaxies. 
	
	A number of authors have obtained constraints on $\sigma_{DM}$ from analytical and semi-analytical arguments.  Strong lensing events are extremely sensitive to the inner profiles and shapes of dark halos in clusters of galaxies, and are thus a powerful probe of SIDM, which has enabled several authors to place constraints on $\sigma_{DM}$ (Miralda-Escud\'{e} 2000; Wyithe, Turner \& Spergel 2001; Meneghetti et al. 2001).  Mo and Mao (2000) determined what value of $\sigma_{DM}$ would produce a correct Tully-Fisher relation for SIDM halos.  Further constraints can be placed on SIDM because of the existence of subhalos in larger halos: heat transfer to a cool subhalo from the hotter halo could evaporate subhalos conflicting with observations if the evaporation were too efficient.  Gnedin and Ostriker (2001) constrained $\sigma_{DM}$ by requiring that the dark halos of elliptical galaxies in clusters survive until the present.  Finally, Ostriker (2000) showed that accretion of SIDM onto seed black holes could produce supermassive black holes in the observed range and with the observed mass scaling $M_{BH}\propto v^{4-5}$ (Magorrian et al. 1998; Ferrarese \& Merritt 2000; Gebhardt et al. 2000; Merritt \& Ferrarese 2001), and he placed an upper limit on $\sigma_{DM}$ so as to avoid the formation of central black holes that are too large.
	
	In this paper we consider the astrophysical consequences of a general velocity dependent cross section, as has been discussed by various authors (Yoshida et al. 2000b; Dav\'{e} et al. 2000; Hogan \& Dalcanton 2000; Gnedin \& Ostriker 2001; Firmani et al. 2001a,2001b; Wyithe et al. 2001; Hui 2001).  We refine the black hole constraint of Ostriker (2000), present new constraints due to other astrophysical considerations, and generalize some of the constraints discussed above for a non-constant cross section.  The ability of SIDM to reproduce the observed properties of the galactic supermassive black hole population is also explored. In \S 2 we determine the black hole mass that will be grown in a given dark halo. We discuss in \S 3 the sensitivity of this result to inner profile flattening. We consider in \S 4 the scaling of black hole mass with halo velocity dispersion. In \S 5 we obtain upper limits on the strength of the dark matter interaction from the observed masses of supermassive black holes.  The effect of the merger history of dark halos is discussed in \S 6.  New constraints on the dark matter interaction are presented in \S 7 and we then summarize and generalize previous ones.  We conclude in \S 8.  Throughout this work we adopt a $\Lambda CDM$ cosmology with  $\Omega_0=0.3$, $\Omega_b=0.045$, $\Omega_{\Lambda}=0.7$, $h_{100}=0.65$, $\sigma_8=0.9$ (see e.g. Bahcall et al. 1999; Eisentstein \& Hu 1999)

\section{SIDM Black Holes}

	A fundamental assumption of our black hole growth scenario is that SIDM dark halos have singular density profiles $\rho \sim r^{-\alpha}$ with $1<\alpha<2$ \emph{immediately after collapse.}  Before discussing black hole formation, we digress briefly to justify this assumption. The essential point is that the collapse occurs on a dynamical time scale ($\sim 10^7 \ \mathrm{yr}$ for galactic halos), whereas the subsequent flattening of the inner profile due to dark matter collisions occurs as the halo evolves secularly on a much longer relaxation timescale ($\sim 10^{10}\ \mathrm{yr} $).  The singular density profiles found in numerical simulations of CDM can been understood with the help of two different analytical pictures. The first is the spherical secondary infall paradigm (Gunn \& Gott 1972; Fillmore \& Goldreich 1985; Bertschinger 1985; Hoffman \& Shaham 1985; Hoffman 1988; Subramanian et al. 1999), whereby a perturbation from the homogeneous cosmological background collapses from the inside out, with subsequent shells of matter collapsing at succesively later times.   Bertschinger (1985) demonstrated that the same singular density profile is to be expected for both collisionless and  \emph{collisional} gases, for the inner parts of the self-similar solutions of the spherical collapse problem.  The alternative picture for explaining cuspy profiles is violent relaxation (Lynden-Bell 1967), the conjecture being that particles will transfer energy to one another as they move through the rapidly varying potential of the collapsing system, resulting in a singular isothermal density profile $\rho \propto r^{-2}$. Whatever the dominant mechanism, both spherical collapse and violent relaxation occur on the dynamical timescale of the halo, which will be much shorter than the timescale for heat transport (via dark matter collisions) which flattens the inner halo (see \S 3).  Kochanek and White (2000) tested this hypothesis by simulating the formation of SIDM dark halos.  They compared the density profiles resulting from the collapse of a top-hat overdensity with and without collisions and found that the final equilibrium profiles are similar and can be fit by cuspy profiles in both cases.  Both the analytical frameworks and Kochanek and White's numerical experiment make it highly unlikely that SIDM dark halo profiles are flat at early times.      


	 To this end, we assume the density profile at early times will be a generalized Navarro-Frenk-White (NFW) (Navarro et al. 1996,1997) or Zhao profile (Zhao 1996)
\be
\rho\left(r\right)=\rho_s\left(\frac{r}{r_s}\right)^{-\alpha} \frac{1}{\left(1+r/r_s\right)^{\epsilon-\alpha}} \ ,\label{NFW}
\ee
so that in the inner regions $r \ll r_s$ the density is given by $\rho\left(r\right)\approx \rho_s\left(\frac{r}{r_s}\right)^{-\alpha}$ with  $1<\alpha<2$ but most likely $\alpha = 1.3 \pm \ 0.2$ (Navarro et al. 1996, 1997; Fukushige \& Makino 1997; Moore et al. 1999; Subramanian et al. 2000; Jing \& Suto 2000; Ghigna et al. 2000; Klypin et al. 2001; Fukushige \& Makino 2001a, 2001b).  Imposing hydrostatic equilibrium for this profile gives the velocity dispersion
\be
v^2\left(r\right) = v_s^2\left(\frac{r}{r_s}\right)^{2-\alpha};  \ \ \  r \ll r_s \ , \label{vs}  \eqnum{2-2a}
\ee
where 
\be
v_s^2 =\mu G\rho_s r_s^2\ ; \ \ \mu\equiv\frac{2\pi}{\left(3-\alpha\right)\left(\alpha-1\right)} \eqnum{2-2b}
\ee
\setcounter{equation}{2}

	Recent high resolution numerical studies of pre-galactic objects indicate that massive stars ($\sim 100 \ M_{\odot}$) form at the centers of progenitor dark matter halos at redshifts $z=15-20$ (Abel, Bryan, \& Norman 2000, 2001).  Subsequent evolution will end in supernovae on Myr timescales, leaving a population of remnant ($\sim 50 \ M_{\odot}$) seed black holes, long before the first galactic mass dark halos have formed (Madau \& Rees 2001).  Therefore, the material that accretes and merges to form galactic mass dark halos will be well seeded with black holes already accreting dark matter on dynamical timescales shorter than the timescale for SIDM collisional evolution.  Following Ostriker (2000), we consider the quasi-spherical accretion of dark matter onto a single seed black hole at the center of a galactic dark matter halo.  Clearly, the assumption of a single seed black hole at the halo center is an over-simplification. However, given the rapid phase of initial growth (see below), one black hole is likely to dominate and will eat or eject the others, so that this complication should not alter our estimates of the final black hole mass.  

	The dark matter is treated as an adiabatic gas, which is valid provided the optical depth diverges for $r<<r_s$ (an assumption we check below).  The density and velocity dispersion can be written
\be
\rho\left(r\right)\approx \left\{ \begin{array}{ll}                                                                             \rho_c\left(\frac{r}{r_c}\right)^{-3/2} & \mbox{$r \ \lesssim \ r_c$} \\  \rho_s\left(\frac{r}{r_s}\right)^{-\alpha} & \mbox{$r_c \ \lesssim r \ \ll r_s$}  \end{array}   \right. ,
\ee
and 
\be
v^2\left(r\right)\approx \left\{ \begin{array}{ll}                                                                             v_c^2\left(\frac{r}{r_c}\right)^{-1} & \mbox{$r \ \lesssim \ r_c$} \\          v_s^2\left(\frac{r}{r_s}\right)^{2-\alpha} & \mbox{$r_c \ \lesssim \ r \ \ll r_s$}  \end{array}   \right.  ,
\ee
where $r_c$ is the ``capture'' or Bondi accretion redius
\be
r_c = \frac{GM_{BH}}{v^2\left(r_c\right)}. \label{rc}
\ee
The quantities $\rho_c$ and $v_c$ can be expressed in terms of $\rho_s$ and $v_s$ by requiring $\rho$ and $v$ be continuous. 

Consider a cross section per unit mass $\sigma_{DM}$, which varies as some power of velocity
\be    
\sigma_{DM}\left(v\right) = \sigma_0\left(\frac{v}{v_0}\right)^{-a}, \label{sigma}
\ee  
where $\sigma_0$, $v_0$, and $a$ are determined by the fundamental physics of the interaction.  Then the optical depth is 
\be
\tau\left(r\right) \approx \left\{ \begin{array}{ll} \tau_c\left(\frac{r}{r_c}\right)^{\left(a-1\right)/2} & \mbox{$r \ < \ r_c$} \\ \tau_s\left(\frac{r}{r_s}\right)^{1-\beta} & \mbox{$r_c \ < \ r \ < \ r_s$} \end{array} \right. ,  \label{tau} \eqnum{2-7a}
\ee
where
\be
\tau_s \equiv \rho_s r_s\sigma_{0}\left(v_s/v_0\right)^{-a}  \ ; \ \tau_c \equiv \rho_c r_c\sigma_{0}\left(v_c/v_0\right)^{-a},\eqnum{2-7b} \label{tau_s}
\ee
and
\be
\beta \equiv a\left(1-\alpha/2\right)+\alpha.  \eqnum{2-7c}
\ee
\setcounter{equation}{7}
For our halo to be optically thick in the inner regions ($\tau \gg 1$ for $r \ll r_c$ and $r \ll r_s$), we must have $a<1$ and $\beta>1$. The latter condition is satisfied for all $a>0$, provided that $1<\alpha<2$ ; however, the former implies a hydrodynamic treatment of the accretion flow is only valid provided the cross section velocity dependence is not too steep.  Hence, the black hole mass we derive below only applies to cross sections with $0<a<1$.

	The characteristic size of our accreting hydrodynamic system during this period of rapid growth is $r_c$ (eqn. \ref{rc}).  The black hole will grow via Bondi-like accretion until the accretion radius is comparable to the mean free path of the dark matter, $\tau\left(r_c\right) \equiv 1$ .  Then a transition will occur to a much slower, diffusively limited growth.  This allows us to obtain a rough estimate of the black hole mass grown during the optically thick phase
\be
M_{BH} \approx \mu \rho_s r_s^3\tau_s^{\frac{3-\alpha}{\beta-1}}. \label{Mbh}
\ee

Thus, the black hole mass will be of order the halo mass, $\rho_s r_s^3$, times the optical depth at the characteristic scale, $\tau_s$, raised to a power which depends on the inner density profile and the velocity dependence of the scattering.  Figure \ref{bhalpha} shows the dependence of the black hole mass on the inner profile slope for a Milky Way size dark halo, $M_{halo} = 3.0 \times 10^{12} \ M_{\odot}$, and several different values of the exponent, $a$, of the cross section velocity dependence.  The black hole mass depends on the combination $\sigma_0 v_0^a$ of the cross section parameters in eqn. (\ref{sigma}).  In Figure \ref{bhalpha} we set $\sigma_1v_{100}^a=1$ where $\sigma_1 \equiv \left(\sigma_0/1 \ \mathrm{cm}^2 \ \mathrm{g}^{-1}\right)$ and $v_{100} \equiv \left(v_0/100 \ \mathrm{km} \ \mathrm{s}^{-1}\right)$.  The characteristic density, radius, and velocity dispersion, $\rho_s$, $r_s$, and $v_s$ have been calculated from the dark halo concentrations using the routine made publicly available by Eke, Navarro, \& Steinmetz (2001) (hereafter ENS), and described in Appendix A.

	Once the accretion radius extends into the optically thin region of the dark halo the fluid approximation is no longer valid.  The subsequent slow growth of the black hole proceeds as dark matter particles are scattered into the loss cone.  This phase of growth was treated by Ostriker (2000) for a velocity independent cross section.  Loss cone accretion can grow black holes comparable to eqn. (\ref{Mbh}) (for a general velocity dependent $\sigma_{DM}$).  However, this assumes a cuspy profile extends into the innermost regions of the halo for a Hubble time, which will not be the case if halos are significantly flattened by heat transfer during that same time interval.  Conversely, accretion from the optically thick region of the halo grows black holes nearly instantaneously in comparison to cosmological timescales, as we will see in the next section.  Since we have neglected the optically thin phase of growth and also any contribution to the mass from baryons, eqn. (\ref{Mbh}) should be regarded as a lower limit when compared to observations.  Finally, it should be noted that the black hole mass calculated here is that grown from a single dark halo.  We will consider the effect of the dark halo merger history on this estimate in \S 6.

\section{Inner Profile Flattening}

	From Figure \ref{bhalpha} it is apparent that the black hole mass in eqn. (\ref{Mbh}) may be very sensitive to the value of the inner profile exponent $\alpha$.   It is thus prudent to determine how physical processes not considered above might alter the inner profile.  In what follows we consider the effect of the accretion flow and heat transfer.
	
	If, to lowest order, we take the accretion to be spherical, then, in the inner regions $r \ll r_c \ll r_s$, gravity dominates over pressure and the particles are in free fall with the radial velocity $u\propto r^{-1/2}$. Then $\dot{M} =  \mathrm{const}$, implies $\rho \propto r^{-3/2}$.  This density profile will interpolate smoothly with the most probable value of the inner profile slope $\alpha = 1.3 \pm 0.2$ at $r\sim r_c$.
	
\begin{figure}
\plotone{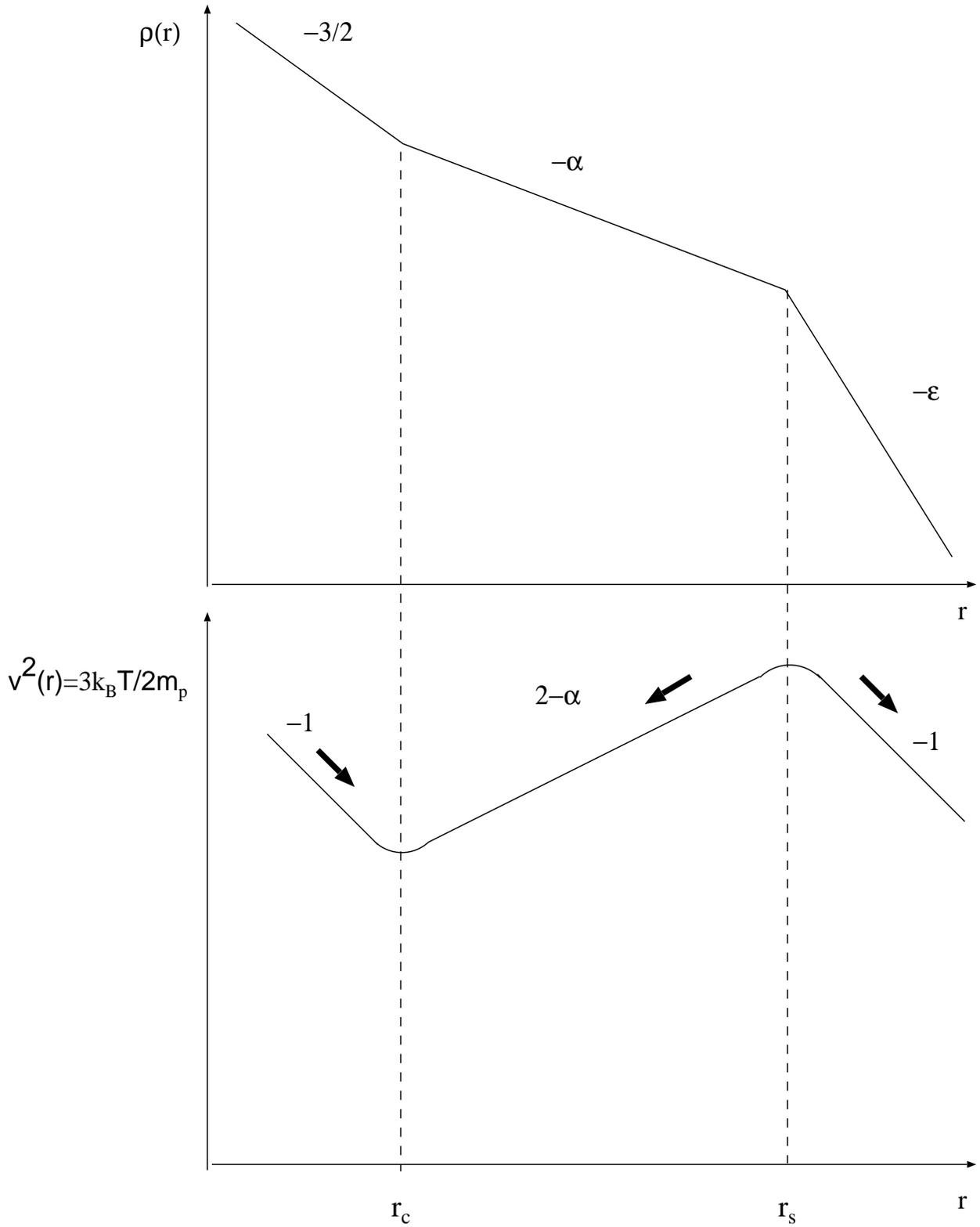}
\caption[SIDM Halo]{Schematic illustration of the density and velocity dispersion (temperature) profiles for SIDM dark halos at early times.  A black hole accretes dark matter from the center of the halo.  Power law slopes are indicated near each segment of the profiles.  Arrows indicate the direction of heat trasfer made possible by the collisional nature of the dark matter.}
\label{cartoon}
\end{figure}

	Now consider the effect of heat transfer.   We have just seen that $u\propto r^{-1/2}$ or $T\propto r^{-1}$ for $r\ll r_c$, so that the inner regions of the flow are dynamically hot.  Also the density profile in eqn. (\ref{NFW}) produces a temperature inversion, so that heat will flow from the outer halo inward flattening the inner profile.  We are thus led to consider how heat flow outward from the accretion flow $r \ll r_c$ and inward from the outer halo $r \gg r_c$ alters the density profile near the temperature minimum at $r \sim r_c$.  Taking note of the fact that the transport behavior differs under optically thick and thin conditions, we consider these cases separately. Figure \ref{cartoon}, presents a cartoon to illustrate the density and temperature profiles and the direction of heat flow for an SIDM dark halo with a black hole accreting from the inner regions.   
	         	 
	For $r<<r_c$ ($\tau \gg 1$), the dark matter behaves like a fluid and hence heat transfer can be described by a diffusion equation  
\be
T\frac{dS}{dt}=\frac{1}{\rho r^2}\frac{\partial}{\partial r}r^2 m_p \kappa\frac{\partial T}{\partial r},\label{therm}
 \ee
where $S$ is the entropy, $m_p$ is the dark matter particle mass, and $T$ is the temperature defined as $\frac{1}{2}k_BT\equiv\frac{1}{2}m_pv^2$.  The coefficient of thermal conductivity for the monatomic gas is 
\be
\kappa\sim c_v\lambda\frac{\rho}{m_p}v=\frac{c_v}{m_p}\left(\frac{k_BT}{\sigma_{DM}m_p}\right)^{1/2}, \label{kappa}
\ee
where $c_v$ is the specific heat per particle and $\lambda$ is the mean free path.  The entropy $S$ relates to the equation of state $P=\frac{\rho}{m_p}k_BT\propto \rho^{\gamma}$. It can be written
\be
T\frac{dS}{dt}=\frac{d}{dt}c_vT+P\frac{d}{dt}\frac{1}{n}=\rho^{\gamma-1}\frac{d}{dt}\frac{c_vT}{\rho^{\gamma-1}}. \label{S}
\ee
Combining eqns. (\ref{therm}), (\ref{kappa}), and (\ref{S}) and  de-dimensionalizing them using the characteristic size $r_c$, density $\rho\left(r_c\right)$, and temperature $T_c\equiv\frac{m_p v^2\left(r_c\right)}{k_B}$, we obtain for the unit of time
\be
t_{thermal}=\tau\left(r_c\right)t_{dyn}\left(r_c\right). \label{ttherm}
\ee
Here, $t_{thermal}$ is the heat conduction timescale, and $t_{dyn}\left(r\right)=r/v\left(r\right)$ is the dynamical time.  Since the dark matter will typically be optically thick at $r_c$ , $t_{thermal} \gg t_{dyn}\left(r_c\right)$.  This condition breaks down in the final stages of black hole growth when the accretion radius $r_c$ approaches the region of the halo of optical depth unity $\tau \sim 1$.  Again, taking the accretion flow to be spherical, one can show that the accretion radius grows (as the mass of the black hole grows) at the local sound speed
\be
\frac{dr_c}{dt}=2\left(\alpha-1\right)v\left(r_c\right). \label{rcdot}
\ee
Then,the speed at which thermal disturbances propagate is 
\be
v_{thermal} \sim \frac{r_c}{t_{thermal}} \sim \frac{v\left(r_c\right)}{\tau\left(r_c\right)} \ll \frac{dr_c}{dt},
\ee
which is much slower than the growth of the accretion radius.  We conclude that heat conduction from the the accretion flow outward cannot alter the density profile fast enough to halt the growth of the black hole, until the accretion radius grows into the region of the halo where $\tau\left(r_c\right) \sim 1$.  

	Now we consider heat conduction inward from the optically thin halo.  Integrating eqn. (\ref{rcdot}) gives $r_c\left(t\right)$; and we can solve for the black hole growth time $t_{BH}$, which is the time for the accretion radius, $r_c$, to grow until $\tau\left(r_c\right)=1$ :
\be
t_{BH}=\tau\left(r_s\right)^{\frac{\alpha}{2\left(\beta-1\right)}}t_{dyn}\left(r_s\right). \label{tbh} 
\ee
For the range of parameters which satisfy all the astrophysical constraints (see \S 7), $\sigma_1v_{100}^a\sim 1$ and $0\lesssim a \lesssim 1$, $\tau_s \lesssim 0.02$ on galactic scales.  Thus, from eqn. (\ref{tbh}) $\left(t_{BH}/t_{dyn}\left(r_s\right)\right) \lesssim 1$, and the dark matter induced phase of black hole growth will be quite rapid, essentially on the timescale of the collapse of the dark matter halo.  From the dwarf galaxy to cluster scale with $\alpha=1.5$, we find $10^5 \  \mathrm{yr} \lesssim t_{BH} \lesssim 10^7 \ \mathrm{yr}$ for $ 0 \leq a \leq 4$.  In particular, for the case $a=1$ discussed in the literature (Yoshida et al. 2000b; Dav\'{e} et al. 2000; Hogan \& Dalcanton 2000; Gnedin \& Ostriker 2001; Firmani et al. 2001a,2001b; Wyithe et al. 2001; Hui 2001), $t_{BH}$ is independent of the scale of the halo and is given by
\be
t_{BH}=\frac{\sigma_0v_0}{\mu G}= 5.7 \times 10^5  \ \mathrm{yr}  
 \left(\frac{\sigma_0}{\mathrm{cm}^2 \ \mathrm{g}^{-1}}\right)\left(\frac{v_0}{100 \ \mathrm{km} \ \mathrm{s}^{-1}}\right).
\ee

	As mentioned above, the post-collapse profile of the dark halo has a temperature inversion.  Ultimately, the conduction of heat inward from the outer halo will flatten the inner profile and a constant density core will form.  This evolution will take place on the relaxation timescale (Kochanek \& White 2000; Burkert 2000; Quinlan 1996, Balberg et al. 2001)
\be 
t_{rel}\left(r_s\right) = \frac{1}{\tau_s}t_{dyn}\left(r_s\right) =   t_{BH}\tau_s^{-\left(1+\frac{\alpha}{2\left(\beta - 1 \right)}\right)} \gg t_{BH}.
\ee
The relaxation time is of order a Hubble time for the optically thin halos considered, so the black holes are effectively grown instantaneously in comparison to the relaxation time and other cosmological timescales. \emph{Although SIDM serves to flatten central density profiles after relaxation timescales, before this occurs $10^6-10^8 \  M_{\odot}$ black holes will have grown in galaxies by optically thick accretion provided their dark halos had a cuspy profile for $\sim 10^6$ yr of their history.}  

\section{Scaling Relations}

\begin{figure}
\plotone{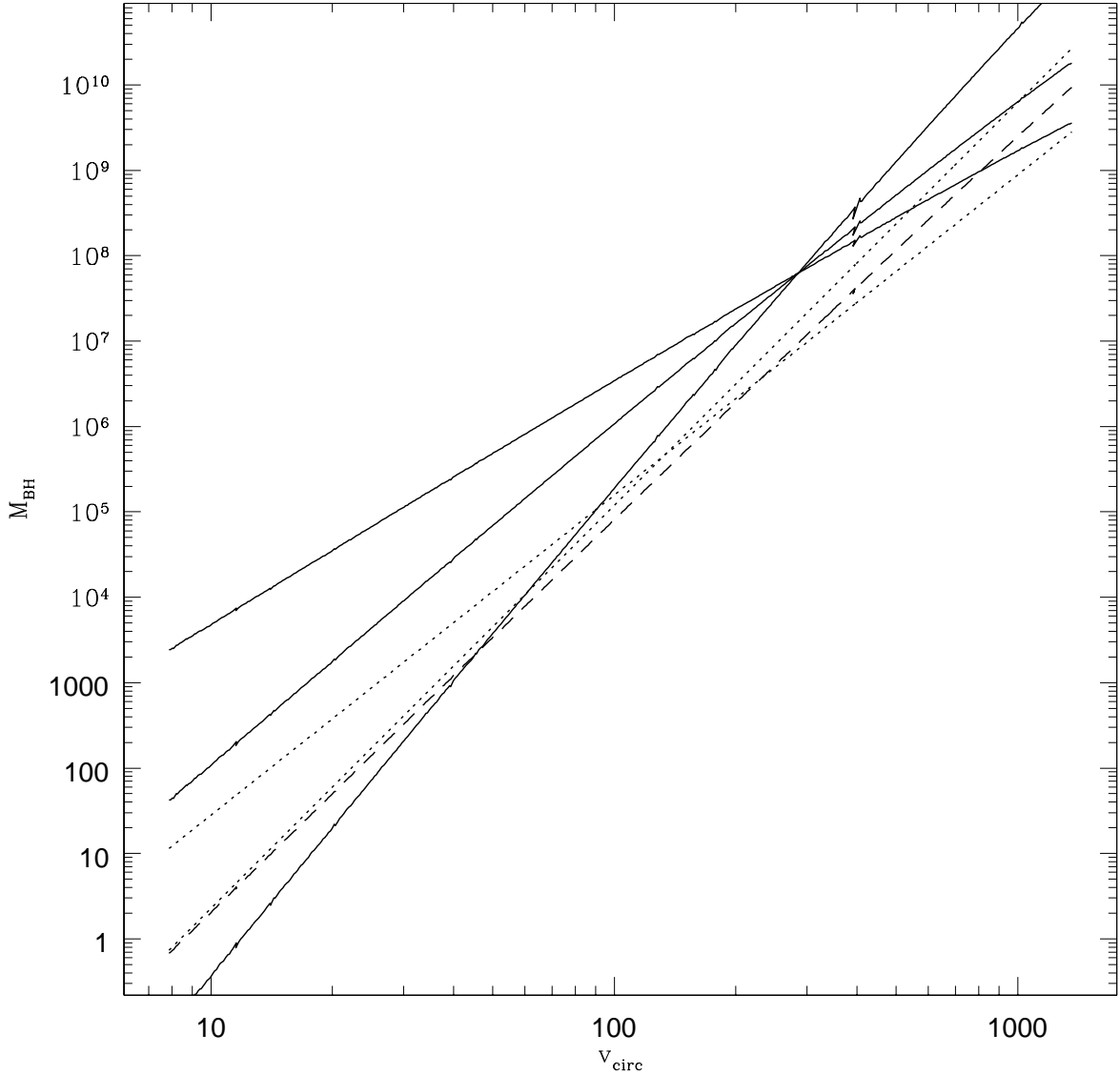}
\caption[Black Hole Mass versus Halo Circular Velocity]{Black hole mass grown from optically thick accretion as a function of dark halo circular velocity for different values of $a$ (the velocity dependence of the cross section) for a Milky Way sized dark halo.  The shallow and steep dotted lines are the observed scaling from Gebhardt et al. (2000) and Merritt \& Ferrarese (2000) respectively.   The solid lines are the $M_{BH}-v$ relation for SIDM with $a=0,0.5$ and $1$, from steepest to shallowest ($a=1$ is the shallowest curve).  The dark matter interaction strength has been set to $\sigma_1v_{100}^a=1$ and the inner profile exponent is $\alpha=1.5$.  The dashed curve is our ``best fit'' model ($\alpha=1.74$, $a=0$, $\sigma_1v_{100}^a=0.02)$}
\label{bhvcirc}
\end{figure}

	Figure \ref{bhvcirc} shows plots of the black hole mass from eqn. (\ref{Mbh}) versus dark halo circular velocity for different values of $a$, where we have taken $\sigma_1 v_{100}^a=1$.  These were generated using the ENS scaling relations described in Appendix A (ENS 2001).  The dashed line is the $M_{BH}-v$ relation determined by Merritt \& Ferrarese (2001), $M_{BH}=1.3 \times 10^8 \ M_{\odot} \left(\sigma_c/200 \ \mathrm{km} \ \mathrm{s}^{-1}\right)^{4.72}$, where $\sigma_c$ is the central velocity dispersion of the bulge.  The dotted line is the shallower $M_{BH}-v$ relation derived by Gebhardt et al. (2000), $M_{BH}=1.2 \times 10^8 \ M_{\odot} \left(\sigma_e/200 \ \mathrm{km} \ \mathrm{s}^{-1}\right)^{3.75}$, where $\sigma_e$ is the luminosity-weighted line-of-sight velocity dispersion within the half-light radius.  Note that, for the sake of comparison to our model, we have extrapolated these observed $M_{BH}-v$ to all dark halo scales, although black holes have only been observed in galactic scale systems.  In order to compare SIDM black holes to observations we must relate the circular velocity of the dark halo to these measured bulge velocity dispersions, $\sigma_c$ and $\sigma_e$ (or equivalently we could relate the mass of the halo to the mass of the bulge).  A determination of these relationships is beyond the scope of this work, but for the sake of comparison we naively assume simple proportionalities, $\sigma_c \propto v_{circ}$ and $\sigma_e \propto v_{circ}$, between the quantities.  The constants are determined from the Milky Way for which $\sigma_c = 100 \ \mathrm{km} \ \mathrm{s}^{-1}$ (Merritt \& Ferrarese 2001)
, $\sigma_e = 75 \ \mathrm{km} \ \mathrm{s}^{-1}$ (Gebhardt et al. 2000), and $v_{circ}=220 \ \mathrm{km} \ \mathrm{s}^{-1}$ (Binney \& Tremaine 1987).

	It is apparent from Figure \ref{bhvcirc} that increasing the exponent $a$ tends to flatten the $M_{BH}-v$ relation for black holes grown from SIDM.  
Ostriker (2000) considered the simplest case, $a=0$, and found $M_{BH} \propto v^{4}$, in satisfactory agreement with observations.  Figure \ref{bhvcirc} also shows our ``best fit'' model to the magnitude and slope of the two observed $M_{BH}-v$ relations, which has $\alpha=1.74$, $\sigma_1v_{100}^a=0.02$, and $a=0$.  Setting $a=0$ reproduces the slope of the observed relations, while the other parameters, $\sigma_1v_{100}^a$ and $\alpha$, have been chosen to be consistent with the discussion in the next section. However, it should be kept in mind that other combinations of $\alpha$ and $\sigma_1v_{100}^a$ would yield a similar $M_{BH}-v$ relation because of a degeneracy in these two quantities.

	The scaling in Figure \ref{bhvcirc} can be understood in a cosmological context as follows.  Suppose a halo of mass $M_s$ collapses to form a virialized object with characteristic size $r_s$, density $\rho_s$, and velocity dispersion $v_s$.  For a power law power spectrum of density fluctuations $P\left(k\right) \propto k^n$ with a spectral index $-3 \leq n \leq 1$, simple scaling arguments from linear theory predict (Peebles 1980; Padmanabhan \& Subramanian 1992; Padmanabhan 1993)
\be
\rho_s \propto v_s^{6\left(n+3\right)/\left(n-1\right)}, \ \ \ r_s \propto v_s^{-2\left(n+5\right)/\left(n-1\right)}.
\ee
Use of these relations in eqn. (\ref{Mbh}) gives the scaling of black hole mass with velocity dispersion.  If $a=0$
\be
M_{BH} \propto v_s^{\frac{4\left(n+2\right)}{\left(n-1\right)}\frac{\left(3-\alpha\right)}{\left(\alpha-1\right)}-\frac{12}{\left(n-1\right)}}.
\ee
On galactic scales the spectral index $n\approx -2$, which gives the approximate scaling $M_{BH} \propto v_s^{4}$, independent of $\alpha$.  For general $a$ and taking $\alpha = 1.5$ and $n=-2$
\be
M_{BH}\propto v_s^{4-\frac{3a}{\left(a/2+1\right)}},
\ee
which roughly explains why the $M_{BH}-v$ relation is flatter for larger $a$.
	
\section{Black Hole Constraints}

	Thus far, we have completely neglected baryons in our discussion of supermassive black holes grown from accretion of dark matter.  Baryons will alter the estimates of $M_{BH}$ in two ways.  First, accretion of baryons will add to the mass of the central black hole so that the black hole mass estimated in \S 2 should be considered a lower limit when compared to observations.  Second and perhaps more significant, as baryons cool and contract they will condense in the inner regions compressing the dark halo and resulting in a steeper inner density profile.  If this compression occurs before the flattening caused by dark matter collisions, then the black holes grown will be significantly larger because of the steep dependence of $M_{BH}$ on $\alpha$ in Figure \ref{bhalpha}.  Observations indicate that in general, bulgeless galaxies do not appear to contain massive black holes comparable with those found in ellipticals of comparable total mass (Richstone et al. 1998; Kormendy 2000; Gebhardt et al. 2001).  In the black hole growth scenario described here, the explanation for this could be that galaxies with inner regions dominated by a stellar bulge, will have compressed their central dark halos thus increasing $\alpha$, whereas, compression will be less significant in bulgeless systems.

	With regards to constraining the dark matter interaction, the question arises as to whether eqn. (\ref{Mbh}) should be compared to bulge or bulgeless systems. In this section, we first obtain a clean constraint by comparing to the bulgeless system M33, where the effects of baryons can be neglected.  In Appendix B, the effect of baryonic infall and compression on the inner profile slope of the dark halo is determined, which is used to obtain another black hole constraint by comparing to the bulge systems that lie on the $M_{BH}-v$ relation.   

\subsection{A Supermassive Black Hole in M33?}

	M33 is a normal low luminosity dark matter dominated spiral (Scd) which lacks a significant bulge.  Its rotation curve is dominated by its dark halo from 3 kpc outward, so that it is relatively easy to disentangle dark and luminous matter and obtain an estimate for the mass of the halo (Persic, Salucci, \& Stel 1996; Corbelli \& Salucci 2000).  Corbelli \& Salucci (2000) analyzed the rotation curve  of M33's H I disk out to 13 disc scale lengths (16 kpc).  From their outermost data points, $v_{circ} \approx 125 \ \mathrm{km} \ \mathrm{s}^{-1}$, which we take as the circular velocity of the dark halo, allowing us to deduce its mass, $M_{halo}\approx 5.1 \times 10^{11} \ M_{\odot}$, from the scaling relations in Appendix A.  M33 also has a notoriously small upper limit on the mass of its central black hole.  Observations from the Space Telescope Imaging Spectrograph (STIS) on the Hubble Space Telescope place an upper limit of $M_{BH} \lesssim 1500 \ M_{\odot}$ (Kormendy et al. 2001; though see Merritt et al. 2001).

	The masses of the dark halo $M_{halo}$ and black hole $M_{BH}$ allow us to obtain a constraint on the dark matter interaction from eqn. (\ref{Mbh}).  Specifically, an upper limit on the quantity $\sigma_1 v_{100}^a$ can be obtained as a function of $a$:
\be
\sigma_1 v_{100}^a \lesssim \left(\frac{M_{BH}}{\mu\rho_s r_s^3}\right)^{\left(\frac{\beta-1}{3-\alpha}\right)}\frac{1}{\rho_s r_s \left(1 \ \mathrm{cm}^2 \ \mathrm{g}^{-1}\right)\left(\frac{v_s}{100 \ \mathrm{km} \ \mathrm{s}^{-1}}\right)^{-a}} = 23.9\left(0.037\right)^{a} \ , \ 2.34\left(0.051\right)^{a}, \ 0.13\left(0.10\right)^{a}  \label{sig33}
\ee
where $\rho_s$, $r_s$, and $v_s$ are determined by $M_{halo}$ from the scaling relations, and we have set $\alpha=1.1$, $1.3$, and $1.5$ respectively to obtain the last equality. The accretion of baryons will increase the black hole mass above our estimate in eqn. (\ref{Mbh}) so that eqn. (\ref{sig33}) is an overestimate and the constraint could be tighter.

	It can be argued that constraining the dark matter interaction based on a single spuriously small black hole is unreasonable considering other effects that could conspire to produce a small black hole in the scenario described in \S 2.  For example, the black hole in M33 could have been ejected in a merger event, or more importantly, the value of $\alpha$ in M33's dark halo could have been small.  Figure \ref{bhalpha} indicates that the black hole mass is extremely sensitive to $\alpha$ for a given physical cross section for $a=0-1$.  Since the real cosmic variance of $\alpha$ is certainly at least $0.1-0.2$ (Subramanian et al. 1999), it is plausible that a smaller than average value of $\alpha$ in the post-collapse dark halo of M33 is responsible for the small black hole in this system.  However, M33 is the best observed local example of a group of several bulgeless galaxies that lack supermassive black holes comparable to those found in bulge systems (Richstone et al. 1998; Kormendy \& Gebhardt 2001):  NGC 4395 (Sm) $M_{BH} \lesssim 8\times 10^4 \ M_{\odot}$ (Fillipenko \& Ho 2001), IC 342 (Scd) $M_{BH} \lesssim 5\times 10^5 \ M_{\odot}$ (Boker et al. 1999), NGC 205 (dE5) $M_{BH} \lesssim 9\times 10^4 \ M_{\odot}$ (Jones et al. 1996).  It is doubtful that cosmic variance in $\alpha$ can account for the small black holes in all of these systems.

\subsection{Black Hole Constraint from Bulge Systems}

	An even tighter constraint on the dark matter interaction can be obtained from bulge systems if compression of the dark halo is allowed for.  In Appendix B we show that condensation of baryons results in a steeper inner density profile with $\alpha^{\prime}=f(\alpha,\xi)$, where $\alpha$ and $\alpha^{\prime}$ are the inner dark matter profile exponents before and after compression respectively, and $\xi$ is the exponent of the inner profile of the total mass density (baryons plus dark matter) after compression.  Specifically, we show that, in \emph{both} the collisionless ($\lambda/r \gg 1$) and highly collisional ($\lambda/r \ll 1$) limits, the final dark matter cusp slope is
\be
\alpha^{\prime}=f\left(\alpha,\xi\right)=\frac{\alpha+3\xi-\alpha\xi}{4-\alpha}. \eqnum{5-2a}
\ee
For a final flat rotation curve, typical of normal bulge dominated galaxies, $\xi=2$, giving the simple result
\be
\alpha^{\prime}=\frac{6-\alpha}{4-\alpha}. \eqnum{5-2b}
\ee
\setcounter{equation}{2}
The most likely CDM value of $\alpha=1.3$, gives $\alpha^{\prime}=1.74$.  The tightest constraint is obtained by comparing to the \emph{smallest} bulge system that lies on the observed $M_{BH}-v$ relation, since roughly speaking the $M_{BH}-v$ scaling for SIDM black holes is as steep as or shallower than that observed.  To this end, we compare to the black hole at the galactic center,$M_{BH}\approx 3.0\times 10^6 \ M_{\odot}$ (Genzel et al. 2000), which is the smallest reliable mass estimate that lies on the observed $M_{BH}-v$ (Ferrarese \& Merritt 2000; Gebhardt et al. 2000; Merritt \& Ferrarese 2001).  Using this value in eqn. (\ref{sig33}), with $r_s$ and $\rho_s$ determined from the scaling relations for $M_{halo} = 3.0 \times 10^{12} \ M_{\odot}$ gives,
 \be
\sigma_1 v_{100}^a \lesssim 0.02\left(0.92\right)^{a}  \label{sigbulge}
\ee
for $\alpha^{\prime}=1.74$.  Our ``best fit'' model ($\alpha=1.7$, $a=0$, and $\sigma_1v_{100}^a=0.02$), mentioned in \S 4, will obviously satisfy the constraint in eqn. (\ref{sigbulge}), since it was chosen to reproduce the magnitude and slope of the observed  $M_{BH}-v$ relation.

\section{Mergers}

	In the black hole formation scenario described in this paper, we have considered the growth of a supermassive black hole from dark matter at the center of a single dark halo from a stellar mass size seed. However, in standard hierarchical structure formation models, massive halos experience multiple mergers during their lifetimes.  This suggests that the mass of the supermassive black hole in a galaxy at $z=0$ could be the sum of smaller black holes grown in its progenitor halos.  From eqn. (\ref{Mbh}), an efficiency for growing black holes in a single dark halo, $\epsilon\left(M_{halo}\right)\equiv\frac{M_{BH}\left(M_{halo}\right)}{M{halo}}$, can be calculated.  For the region of parameter space which satisfies all constraints (see below), $0\lesssim a \lesssim 1$, we find that $\epsilon \propto M_{halo}^p$, where $-0.1\lesssim p \lesssim 1.4$ for $\alpha\approx 1.3$ ($-0.1\lesssim p \lesssim 0.5$ for  $\alpha\approx 1.7$), and larger values of $p$ correspond to smaller values of $a$.  Thus, the efficiency is roughly a constant for $a=1$, and varies as a small positive power of halo mass for $a<1$. While a detailed calculation of halo merger dynamics is beyond the scope of this work, it should be noted that because $\epsilon\left(M_{halo}\right)$ is a weak function of $M_{halo}$ for the models considered, we make little error in simply using the present dark halo to calculate the black hole mass. 


	The fact that our scenario populates dark halos at high redshift with supermassive black holes has other interesting astrophysical consequences.  Consider our ``best fit''  model with $\alpha=1.7$, $a=0$, and $\sigma_1v_{100}^a=0.02$.  For this set of parameters the efficiency for growing black holes in a given dark halo is approximately a constant, $\epsilon \propto M_{halo}^{0.5}$, so that the black hole in each dark halo will be proportional to the mass of the halo.  The collisional cross section for this model is so small that collisional relaxation will not ameliorate the cuspy halo problem, as the relaxation timescale will be much longer than the Hubble time.  However, N-body simulations of mergers of galaxies containing supermassive black holes have demonstrated that mergers between galaxies with steep power-law density cusps produce remnants with shallower power-law cusps, because the formation of a black-hole binary transfers energy to the halo, lowering the central density (Merritt \& Cruz 2001; Milosavljevi\'{c} \& Merritt 2001).  Consider then, the early generations of progenitor dark halos at high redshift that grew black holes by accretion of SIDM.  This generation might correspond to the epochs at which the first massive stars collapsed to form seed black holes (Madau \& Rees 2001; Abel et al. 2000, 2001).  In accordance with our calculation in Appendix B, we expect an inner profile cusp of $\alpha \approx 1.7$, in these systems, and black holes will be grown in each dark halo with an efficiency $\epsilon$.  Successive mergers between progenitor halos harboring supermassive black holes will result in a milder density cusp in each remnant, so that parent halos at $z=0$ will have significantly shallower density profiles, thus providing a possible solution to the cuspy halo problem.  

	As inner profile cusps become shallower at late times, there will come a point at which black holes only grow by merging: accretion of SIDM will be negligible because of the steep dependence of $M_{BH}$ on $\alpha$.  Furthermore, because the efficiency, $\epsilon$, is roughly constant, the black hole masses will scale in proportion to the masses of their host halos.  Then, as pointed out by Haehnelt and Kauffmann (2000), the scaling $M_{halo} \propto v_s^{-\frac{12}{\left(n-1\right)}}$ for power law power spectra (see \S 4), gives $M_{BH} \propto v_s^{4}$ at galaxy scales where $n\approx -2$.  Given this scaling initially, mergers between galaxies will move black holes along the observed $M_{BH}-v$ relation.  Finally, it should be noted that the dark matter cusp after compression $\alpha^{\prime}$, will have much less scatter than the initial profile, since a range of $1<\alpha<2$ is compressed to $1.67<\alpha^{\prime}<2$.  The corresponding black hole mass will also be less sensitive to $\alpha^{\prime}$ as the curves in Figure \ref{bhalpha} are flatter for larger $\alpha$. It follows that the efficiency $\epsilon$ and $M_{BH}-v$ scaling are determined by a set of \emph{fixed} parameters: the fundamental physics of the dark matter interaction determines $\sigma_1v_{100}^a$ and $a$ and the universal (initial) profile of dark halos with $\alpha=1.3\pm 0.2$ coupled with the adiabatic collapse of baryons (see Appendix B) determines $\alpha^{\prime}\approx 1.7-1.8$. 

	Accordingly, the black hole growth scenario with the parameters of our ``best fit'' model has the following three consequences: 1) The observed magnitude and scaling of the observed $M_{BH}-v$ relation (at $z=0$) is reproduced. 2) The lack of comparable supermassive black holes in bulgeless galaxies like M33 is explained 3) It provides a possible solution to the cuspy halo problem via mergers.         
	
	This scenario becomes all the more plausible when one realizes that massive black holes need only grow in a few percent of progenitor dark halos.  Menou, Haiman, \& Narayanan (2001), have shown that the presence of central massive black holes at the centers of nearly all nearby galaxies can arise from their merger history, even if only a small fraction ($\sim 3 \times 10^{-2}$) of the progenitor halos harbored black holes at high redshift.  Hence, even if there were significant scatter in the final inner profile slope $\alpha^{\prime}$, which would yield a corresponding scatter in black hole mass (because of the steep dependence of $M_{BH}$ on $\alpha$), our model would still have the aforementioned consequences as a result of the merger history of dark halos.

\section{Constraints from Other Observations}

	The evolution implied by the collisional nature of dark matter must not conflict with observations of dark halos over nearly three orders of magnitude in halo circular velocity.  In this section we discuss four constraints on the dark matter interaction obtained at different halo mass scales.  First, a minimum cross section per unit mass is required to solve the cuspy halo problem on dwarf galaxy and low surface brightness (LSB) galaxy scales, below which SIDM interpolates smoothly with CDM and is astrophysically uninteresting.  Second, all dark halos observed today must not have undergone core collapse by the present, giving a upper limit or core collapse constraint.  Third,  as pointed out by Gnedin and Ostriker (2001), an upper limit on the dark matter interaction can be obtained by requiring that galactic subhalos survive until the present in hotter cluster environments. Finally, an observational upper limit on the core radius of a cluster of galaxies (Arabadjis et al. 2001) also gives a strong constraint on the dark matter interaction.

\subsection{Uninteresting Limit and Core Collapse Constraint}

	In the regime where SIDM halos are optically thin, the dynamics resembles two body relaxation in globular clusters. Heat will flow inward from the outer halo due to the temperature inversion implied by the post-collapse profile (see Figure \ref{cartoon}).  A flattened core will develop and grow outward, with the central density falling and the velocity rising.  Once the temperature inversion is gone, expansion halts.  The direction of heat flow reverses and core collapse begins.  The state of evolution of a dark halo will be determined by the ratio of the relaxation time at its characteristic scale to its age.  For example, if core collapse begins after a number $C_{1}$ relaxation times, then all halos observed today must have
\be
C_{1} \ t_{rel}\left(r_s\right) = \frac{C_{1} \ t_{dyn}\left(r_s\right)}{\tau_s} \gtrsim t_{H} - t_f, \label{cc1} \eqnum{7-1a}
\ee          
where $\tau_s$ is given by eqn. (\ref{tau_s}), $t_{H}$ is the age of the universe, and $t_{f}$ is the halo formation time. Furthermore, if a number $C_{2}$ relaxation times must pass before the central density cusp is significantly flattened, then SIDM will not resolve the cuspy halo problem unless
\be
C_{2} \ t_{rel}\left(r_s\right)\lesssim t_{H} - t_f. \label{cc2} \eqnum{7-1b}
\ee 
Or,
\be
\frac{t_{H} - t_f}{C_1} \lesssim t_{rel}\lesssim \frac{t_{H} - t_f}{C_2}, \label{cc3} \eqnum{7-1c}
\ee
\setcounter{equation}{1} 
a constraint which must be satisfied by all non-cuspy halos.  The constants $C_{1}$ and $C_{2}$ can be calibrated against simulations. 
	
	The formation times in eqn. (\ref{cc3}) are calculated by taking the median of the formation time distribution of Lacey \& Cole (1993), which is based on the extended Press-Schecter formalism (Press \& Schecter 1974; Bond et al. 1991).  The formation time of a halo of present mass $M$ is defined as the time when a parent halo appeared which had half or more its mass.  This particular definition of formation somewhat remedies the neglect of accretion and merging in eqns. (\ref{cc1}a) and (\ref{cc2}b), since the countdown to core collapse starts only after the last major merger in a halos formation history.  Table 1 lists formation times and redshifts of dark halos from the dwarf to cluster scale.  These have been calculated using the analytic fitting formulae in  Appendix B \& C of Kitayama \& Suto (1996) for the CDM power spectrum (Bardeen et al. 1986) and formation time distribution (Lacey \& Cole 1993).

\begin{figure}
\plotone{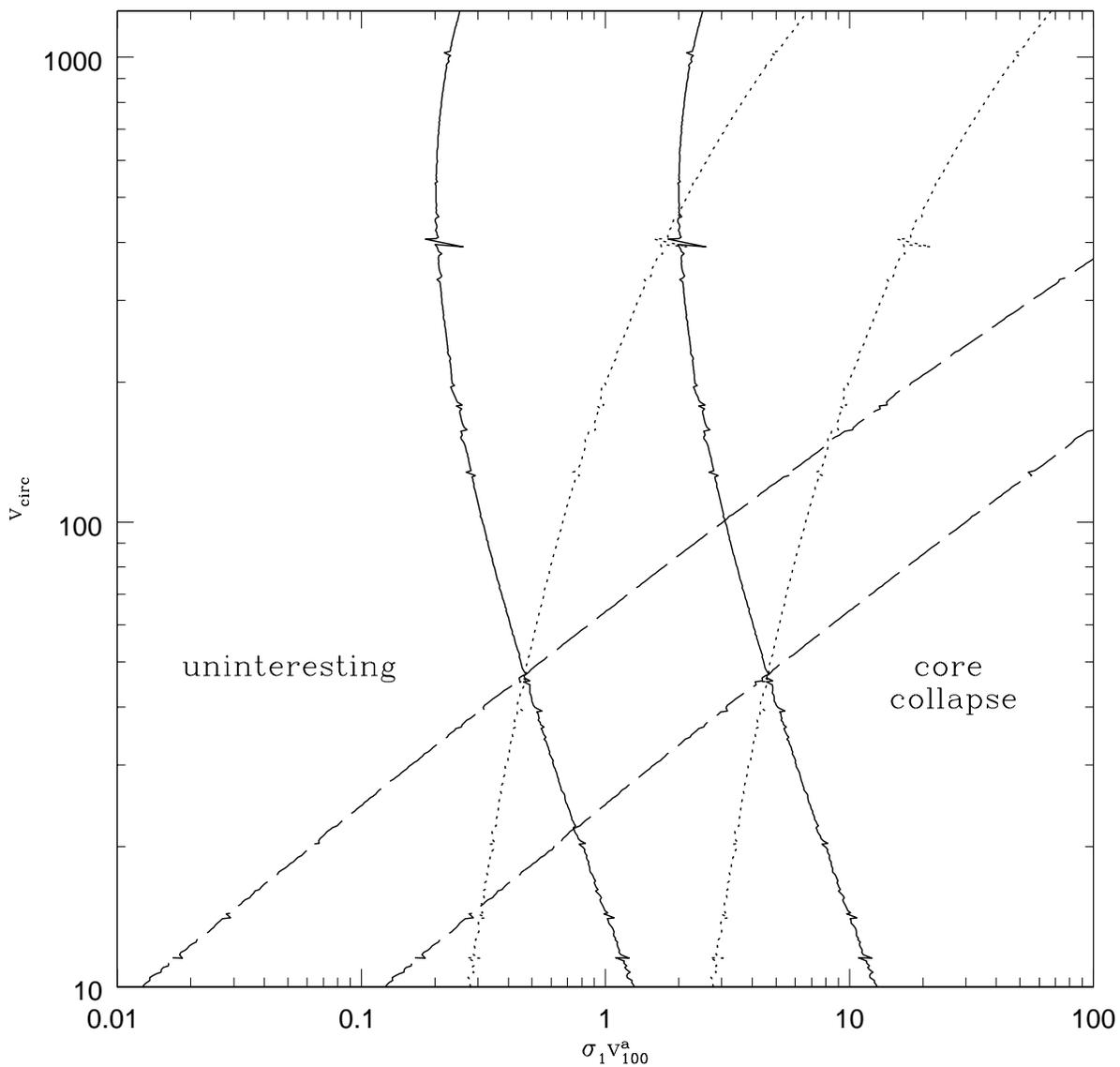}
\caption[Plot of Equation $\ref{cc}$]{Plot of the inequalities in eqns. (\ref{cc1}a) and (\ref{cc2}b) for $C_2=0.48$  and $C_1=4.76$.  The leftmost and rightmost curves correspond to eqn. (\ref{cc1}a) and (\ref{cc2}b) respectively.  Each set of parallel curves corresponds a different value of $a$ and the region between each pair satisfies both inequalities.  Solid, dotted, and dashed lines refer to  $a=0,1,$ and $3$ respectively.}
\label{sigvcirc}
\end{figure}

	Note that the relaxation time and age of a halo will depend on the scale under consideration, so that systems at different scales will be at different stages of evolution.  Specifically, the velocity dependence of the cross section in eqn. (\ref{sigma}) implies lower collision rates in larger systems, so that, for example, dwarf galaxies will be more relaxed than clusters.  Figure \ref{sigvcirc} plots the inequalities in eqns. (\ref{cc1}a) and (\ref{cc2}b) in the $\sigma_1v_{100}^a$ - $v_{circ}$ plane, where $v_{circ}$ is the circular velocity of the dark halo,  for several values of $a$ and values of $C_1$ and $C_2$ taken from simulations. The leftmost and rightmost curves correspond to eqn. (\ref{cc2}b) and (\ref{cc1}a) respectively.  Each set of parallel curves corresponds to a different value of $a$ and the region between each set of curves satisfies both inequalities.  To the left of the leftmost curve inequality (\ref{cc2}b) is violated: $\sigma_1v_{100}^a$ is too small and SIDM halos are indistinguishable from CDM halos.  To the right of the rightmost inequality (\ref{cc1}a) is violated: $\sigma_1v_{100}^a$ is too large and dark halos have core collapsed.  At fixed $v_{circ}$ the region between the curves indicates the range of $\sigma_1v_{100}^a$ for which halos at that scale will be significantly evolved due to collisions.  Or for fixed $\sigma_1v_{100}^a$, it indicates the range of scales that will have undergone the desired amount of evolution.

  	From the curves in Figure \ref{sigvcirc} for $a=0$, a clear but somewhat surprising result emerges: the quantity $\frac{t_{rel}}{t_H-t_f}$ is nearly a constant over three orders of magnitude in circular velocity.  A constant cross section thus implies dwarf galaxies will be just as relaxed as clusters.  However, observations indicate that dwarf galaxies have larger cores in proportion to their characteristic size than do clusters.  There is thus a scaling problem for a constant cross section as pointed out by Dav\'{e} et al. (2000) and Yoshida et al. (2000b).  A velocity dependent cross section obviously remedies this problem since collisions will be more frequent in lower velocity environments and smaller systems will thus be more evolved.  However it is apparent from Figure \ref{sigvcirc} that if the velocity dependence is too steep ($a$ is too large) only a narrow window of scales will have evolved significantly and all larger scales will be identical to CDM halos.   We can obtain a rough estimate of the value of $a$ that will reproduce observations of core sizes on both dwarf galaxy and cluster scales as follows.  Kochanek \& White's (2000) simulations indicate that the core radius grows linearly during the expansion phase of halo evolution, so that $r_{c} \propto \left(r_s\frac{t_H-t_f}{t_{rel}}\right)$.  Then the ratio of cluster to dwarf core radii, $r_{c,cl}/r_{c,dw}$, will depend only on $a$, or solving for $a$
\be
a=\frac{Log\left[\frac{t_H-t_{f,cl}}{t_H-t_{f,dw}}\frac{\left(\rho_{s}r_{s}v_{s}\right)_{cl}}{\left(\rho_{s}r_{s}v_{s}\right)_{dw}}\frac{r_{c,dw}}{r_{c,cl}}\right]}{Log\left(v_{s,cl}/v_{s,dw}\right)}. \label{core}
\ee
Consider DDO 154, a dwarf galaxy which has a core radius $r_{c,dw} \approx 2.5$ kpc and maximum rotational velocity $v_{max}\approx 47 \ \mathrm{km} \ \mathrm{s}^{-1}$ (Carnigan \& Purton 1998), which can be identified with the circular velocity of the dark halo.  The lensing cluster EMSS 1358+6245 has a core radius $r_{c,cl} \lesssim 40$ kpc and an inferred halo mass of $M_{halo} \approx 4\times 10^{14} \ M_{\odot}$ (Arabadjis, Bautz, \& Garmire).  Using the scaling relations in Appendix A to calculate $\rho_s$, $r_s$, and $v_s$, eqn. (\ref{core}) gives $a\approx 0.6$. 

\begin{figure}
\plotone{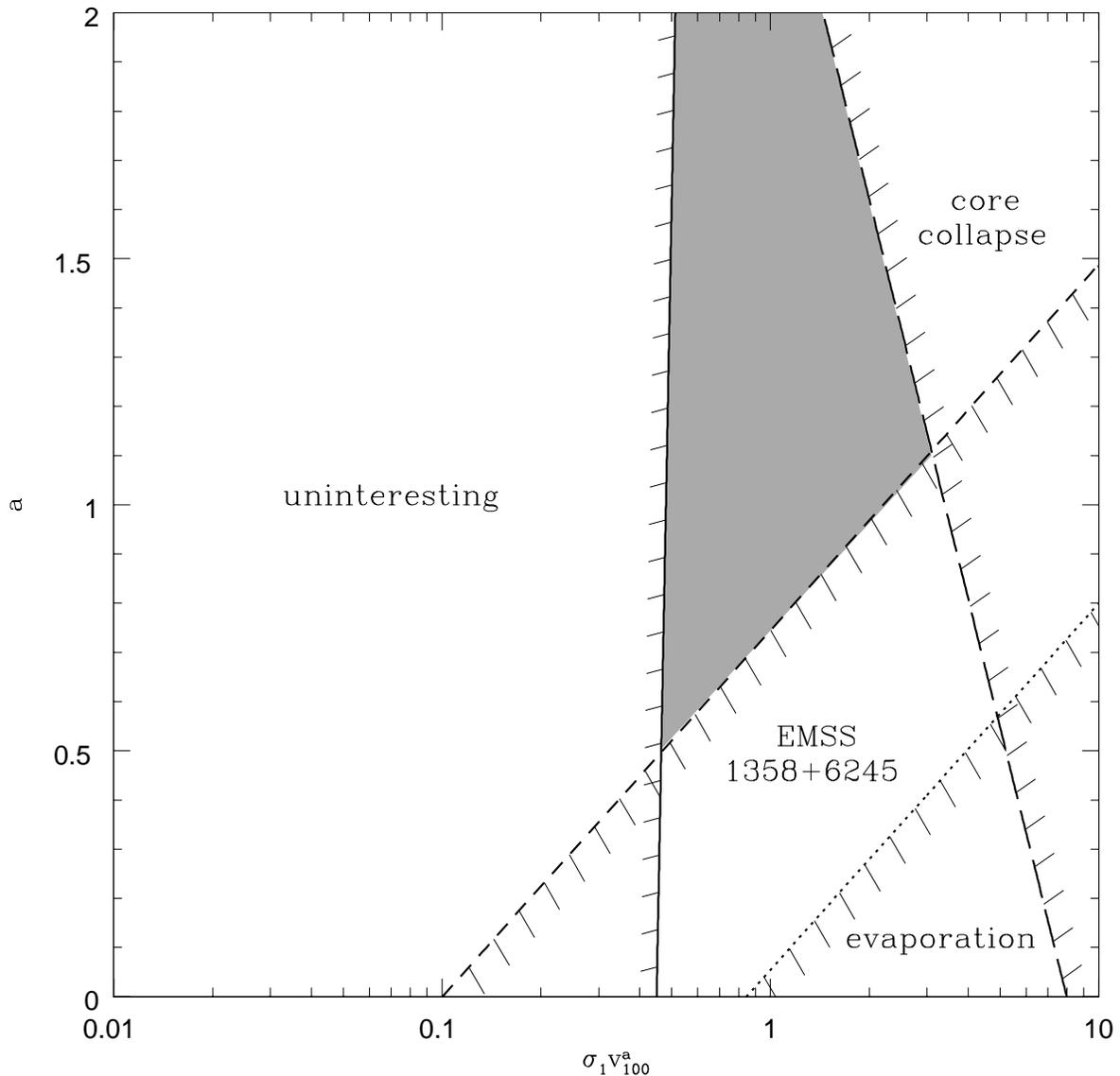}
\caption[Four Constraints on The Dark Matter Interaction]{Constraints on the dark matter interaction plotted in the $\sigma_1v_{100}^a \ - a$ plane.  The solid line is the uninteresting limit, corresponding to the minimum cross section required to flatten dwarf and LSB galaxies.  The dashed curve is the core collapse constraint, obtained by requiring that $20 \ \mathrm{km} \ \mathrm{s}^{-1}$ dark halos have yet to undergo core collapse.   The dotted line is the upper limit adapted from the evaporation constraint of Gnedin \& Ostriker (2001).  The constraint obtained from Chandra observations of cluster EMSS 1358+6245 (Arabdjis et al. 2001) is the short dashed line. Hash marks indicate the region excluded by each constraint. The shaded region is consistent with all four constraints.}
\label{svsa}
\end{figure}

	The curves in Figure \ref{sigvcirc} can be translated into constraints on the dark matter interaction by fixing $v_{circ}$ and thus specifying a physical scale.  For example, since smaller mass halos will always be more evolved relative to larger halos, an upper limit can be obtained in the $\sigma_1v_{100}^a-a$ plane by requiring that the smallest dark halos observed today have yet to undergo core collapse.  We designate $v_{circ}= 20 \ \mathrm{km} \ \mathrm{s}^{-1}$ halos as the smallest observed today and generate a core collapse constraint which is plotted in Figure \ref{svsa}.  Similarly, a lower limit on the dark matter interaction is obtained by requiring that $v_{circ}=50 \ \mathrm{km} \ \mathrm{s}^{-1}$ halos, corresponding to dwarf and LSB galaxies, have undergone significant evolution by the present.  Below this limit, shown in Figure \ref{svsa}, SIDM effectively becomes CDM and is astrophysically uninteresting.
 
	These constraints depend upon the constants $C_1$ and $C_2$ in eqns. (\ref{cc1}a) and (\ref{cc2}b) respectively, which, strictly speaking could be functions of $a$ since the dynamics could differ slightly for different cross section velocity dependences.  However, accurate simulations have only been carried out for a constant cross section so we take them as constants and calibrate to simulations for $a=0$.  Quinlan's (1996) Fokker-Planck simulations indicate that density profiles containing temperature inversions enter a core collapse phase after approximately 5 half-mass relaxation times.  Burkert's (2000) N-body simulations roughly agree with this result, but Kochanek \& White (2000) saw much faster evolution towards collapse.  Balberg et al. (2001) showed that the core collapse time of an SIDM halo is 290 central relaxation times, however this is based on the assumption that the halos have flat inner profiles at all times, which, as per our discussion at the beginning of \S 2, is highly unlikely to hold.   We took $C_1=4.76$ in agreement with the core collapse timescale determined by Kochanek \& White (2000), converted to our units. Simulations indicate that a minimum value of $\sigma_{DM} = 0.45 \ \mathrm{cm}^2 \ \mathrm{g}^{-1}$ is required to flatten dwarf galaxy and LSB galaxy halos (Dav\'{e} et al. 2000; Wandelt et al. 2000).  Plugging this cross section into eqn. (\ref{cc2}b) at the scale $v_{circ}=50 \ \mathrm{km} \ \mathrm{s}^{-1}$ with $a=0$ gives $C_2=0.48$.

\subsection{Evaporation Constraint}

	The existence of subhalos in larger halos further constrains the dark matter interaction.  For galactic subhalos in clusters, heat transfer to a cool subhalo from the hot cluster environment could evaporate the galactic halo if heat transfer were too efficient. This would violate the fundamental plane relations in conflict with observations.   Gnedin and Ostriker (2001) excluded a range of dark matter cross section per unit mass by requiring that galactic subhalos survive until the present epoch.  They considered both optically thick and thin halos and obtained lower and upper limits respectively on the dark matter interaction strength.  Both upper and lower limits can be obtained because of the non-monotonic dependence of the heat transfer on the interaction cross section.  We concern ourselves only with the upper limit obtained in the optically thin regime, since, as mentioned previously cross sections large enough to give optically thick halos can be excluded on other grounds (Dav\'{e} et al. 2000; Wandelt et al. 2000).  To obtain their scattering regime constraint, Gnedin and Ostriker required the cluster relaxation time evaluated at the position of the galactic halo be longer than the age of the cluster, which they approximate as the Hubble time.  A typical galactic halo will be located at the scale radius of the cluster $r_s$, so that the evaporation constraint is just given by eqn. (\ref{cc1}a).  As a conservative estimate, we take the constant $C_1=2$ for evaporation.  This constraint is plotted in Figure \ref{svsa} for a cluster like EMSS 1358+6245 with a halo mass $M_{halo} \approx 4\times 10^{14} \ M_{\odot}$ (Arabdjis et al. 2001).

\subsection{Cluster Core Constraint: Cluster EMSS 1358+6245}

	Recently Arabadjis, Bautz and Garmire (2001) used high resolution Chandra observations of the lensing cluster EMSS 1358+6245 to constrain the dark matter interaction strength.  Specifically, an upper limit of $42 \ \mathrm{kpc}$ on the size of any constant density core in the $M_{halo} \approx 4\times 10^{14} \ M_{\odot}$ cluster was obtained from their X-ray determination of the cluster mass profile.  This upper limit was then compared to Yoshida et al.'s (2000b) numerical simulation of cluster size SIDM halo, where a comparable mass cluster had a core $\sim 40 \ \mathrm{kpc}$ for $\sigma_{0}=0.1$ ($a=0$). This allowed Arabadjis et al. (2001) to place an upper limit of $\sigma_{0} \lesssim 0.1$ ($a=0$) on the scattering cross section.  In the context of the discussion in this section, Arabadjis et al.'s constraint is equivalent to evaluating eqn. (7-1a) for $M_{halo} \approx 4\times 10^{14} \ M_{\odot}$, $\sigma_{0}=0.1$, and $a=0$, thus determining the number of relaxation times ($C_1$) it takes for the cluster to develop a $\sim 40 \ \mathrm{kpc}$ core.  This equation is used to plot the lensing cluster constraint in Figure \ref{svsa}.

\section{Conclusion}

	The shaded area in Figure \ref{svsa} indicates the region of the $\sigma_1v_{100}^a-a$ plane consistent with the four constraints discussed in the previous section.  A window of possible dark matter interactions with $0.5 \lesssim a \lesssim 3 $ and $0.5\lesssim \sigma_1v_{100}^a \lesssim 5$ appears to satisfy the uninteresting, core collapse, evaporation, and cluster EMSS 1358+6245 constraints, which is broadly consistent with the results of numerical simulations (for the regions of parameter space that have been simulated). 

\begin{figure}
\plotone{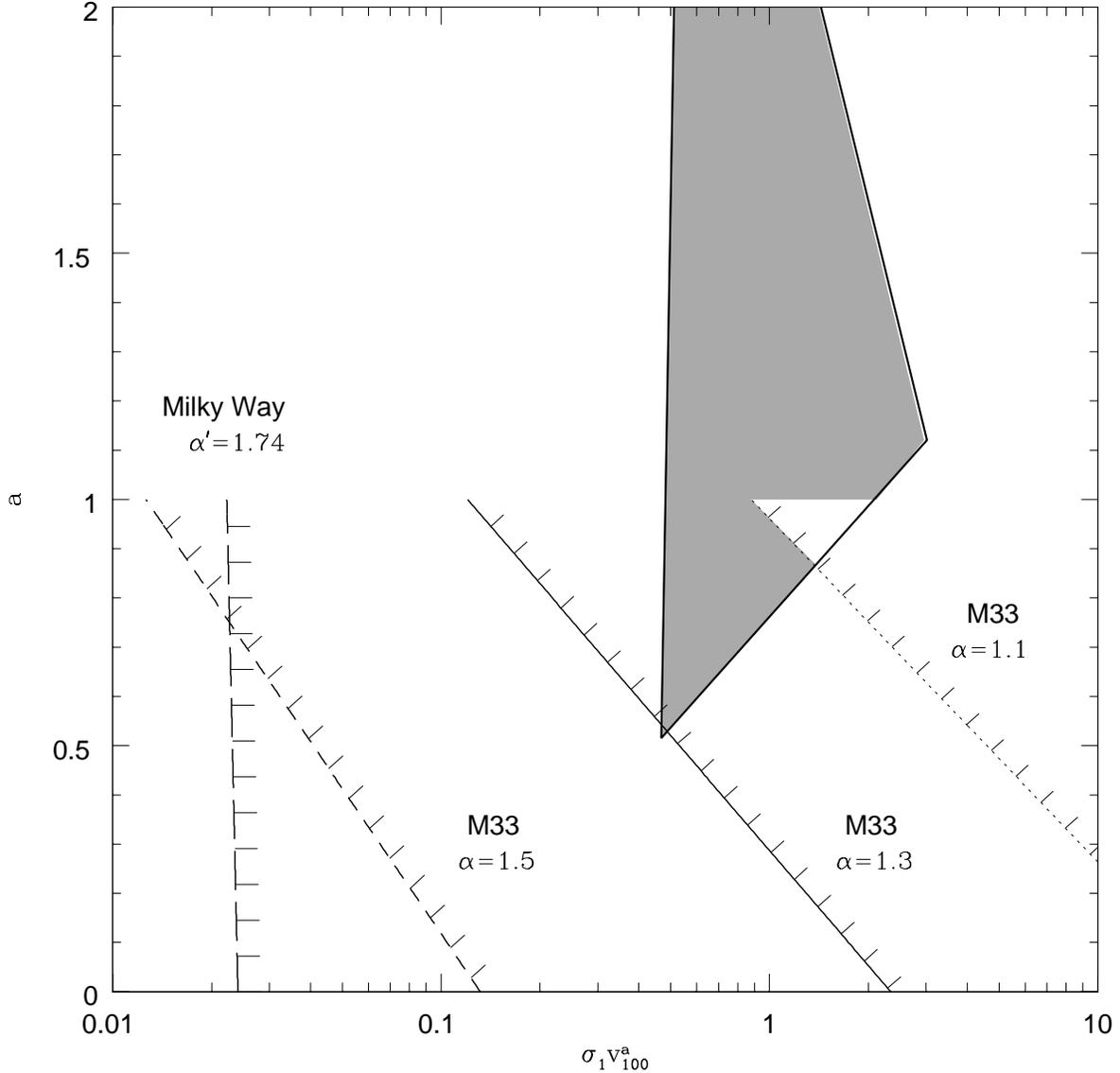}
\caption[Black Hole Constraints on The Dark Matter Interaction]{Black hole constraints on the dark matter interaction plotted with the available region of parameter space for SIDM.  Dotted, solid, and  short dashed lines are the black hole constraints obtained by comparison to the bulgeless galaxy M33 for $\alpha = 1.1$, $1.3$, and $1.5$ respectively. Hash marks indicate the region excluded by each constraint. The large triangle is the region of parameter space allowed by the constraints discussed in \S 6 and shown in Figure \ref{svsa}.  Shading indicates the region which is consistent with the constraints from \S 6 and the M33 constraint with $\alpha=1.1$. The long dashed curve is the black hole constraint obtained by comparing to the Milky Way after taking into account the adiabatic compression of the dark halo by baryons, which changes the inner profile slope to $\alpha^{\prime}=1.74$. The region to the left of this curve is permitted.}  
\label{conc}
\end{figure}

	In Figure \ref{conc} the permitted region of parameter space is plotted with the black hole constraints from eqns. (\ref{sig33}) and (\ref{sigbulge}) for both bulgeless (M33) and bulge (Milky Way) systems.  The bulgeless constraint is plotted for three plausible values of the inner profile slope (without baryonic compression) $\alpha=1.1$, $1.3$, and $1.5$ from right to left.  The region to the left of each respective curve is permitted by the constraint.  The bulge constraint is plotted for a final profile cusp of $\alpha^{\prime}=1.74$ as determined in Appendix B for an initial profile cusp with a most probable value of $\alpha=1.3$ (even if $\alpha=1.0$, the NFW value, the final cusp is $\alpha^{\prime}=1.67$).  The region to the left of the nearly vertical curve is permitted.  

	From the bulgeless constraint in Figure \ref{conc}, it is apparent that if the post-collapse density profile of the dark halo of M33 had $\alpha \gtrsim 1.3$, no region of the parameter space of dark matter interactions exists that is consistent with both the black hole in M33 and constraints from other observations.  This would exclude SIDM as a possible solution to the problems with CDM on subgalactic scales.  If $\alpha=1.3$, a tiny region is still allowed with $a \approx 0.5$ and $\sigma_{1}v_{100}^a \approx 0.5$, corresponding to lower vertex of the triangle in Figure \ref{conc}.  A much larger region of parameter space would be available for $\alpha \lesssim 1.1$. The gray shading in the figure indicates the region of parameter space allowed if $\alpha=1.1$ for M33; however, N-body simulations of dark matter clustering indicate that cusps this mild are unlikely. Apparently, the cross section velocity dependence $a=1$ discussed extensively in the literature can be excluded. Only detailed N-body simulations can determine whether a dark matter cross section consistent with these regions of parameter space can reproduce the observed structure of dark halos on all scales.  With regards to black holes, the parameters quoted above with $\alpha=1.3$, $a\approx0.5$, $\sigma_{1}v_{100}^a \approx 0.5$, would account for $\lesssim 1\%$ of the supermassive black hole mass at the centers of galaxies.  This is roughly the ratio of the observed upper limit on the black hole in M33, $M_{BH} \lesssim 1500 \ M_{\odot}$, to the $M_{BH} \sim 5 \times 10^{5} \ M_{\odot}$ that M33 would have if it lay on the observed $M_{BH}-v$ relation.  This correspondence arises of course because we have forced these parameters to be consistent with the black hole constraint from M33.  Although this fraction is small, the resulting black hole masses (typically $\gtrsim 10^4$ for normal galaxies) are far more massive than stellar mass black holes, and are appropriate seeds for rapid baryonic growth.

	The constraint from bulge systems, which takes into account the probable compression of the dark halo by baryons, is much tighter than the bulgeless constraint, effectively requiring $\sigma_1v_{100}^a \lesssim 0.02$ for $a<1$.  Interaction strengths consistent with this constraint are too small to produce significant collisional evolution of dark halos, since $t_{rel} \gg t_H-t_f$, so that in the absence of other dynamical processes, SIDM behaves effectively as CDM, failing to remedy CDM's small scale problems.  Nevertheless, the interaction parameters in the range permitted by the bulge constraint, would still seed dark halos with supermassive black holes at high redshift.  As mentioned in \S 6, SIDM could then provide an indirect solution to the dark halo cusp problem, since successive mergers of dark halos harboring supermassive black holes at high redshift would result in remnants with flatter inner density profiles.

	Balberg \& Shapiro (2001) have considered an alternative supermassive black hole formation scenario, whereby supermassive black holes are formed after SIDM halos undergo gravothermal collapse. However, for the range of dark matter cross sections permitted by the constraints in Figure \ref{conc}, our calculations of the core collaspe timescale in \S 7 and depicted in Figure \ref{sigvcirc}, indicate that core collapse is highly unlikely for dark halos with circular velocities between 10-1000 km/s.

	In sum, the self interacting dark matter scenario remains of great interest. If the dark matter interaction is a steep function of velocity ($a>1$), a broad range of parameter space with $1.0 \lesssim a \lesssim 3 $ and $0.5\lesssim \sigma_1v_{100}^a \lesssim 5$, is available. Though, as mentioned previously, a steep velocity dependence necessarily implies only a small range of dark halos masses will show significant collisional evolution (see Figure \ref{sigvcirc}).  For shallower velocity dependences ($a<1$), the range of astrophysically permitted interaction strengths imply heat conduction driven flattening of dark halo profiles is unimportant. However, the early and efficient growth of massive black holes at the centers of dark halos is a natural consequence of the theory, which in turn, through succesive mergers, leads to the observed density profiles in galaxies, produces black holes with the appropriate mass and mass scaling, and explains the lack of massive black holes at the centers of bulgeless galaxies like M33.

\bigskip
\acknowledgements
	We would like to thank John Arabadjis, Romeel Dav\'{e}, Oleg Gnedin, Jeremy Goodman, Martin Rees, Uros Seljak, David Spergel, Paul Steinhardt, and Scott Tremaine for helpful discussions.  Special thanks to Martin Rees for pointing out an error in an early version of the manuscript.   

	J. F. Hennawi was supported in part by a grant from the Paul \& Daisy Soros Fellowship for New Americans.  The program is not responsible for the views expressed.

\appendix

\section{ENS SCALING RELATIONS}

	The post collapse density profile of an SIDM dark halo will be identical to that of a CDM halos.  High resolution numerical simulations (Navarro et al. 1996,1997) have shown that these profiles may be fitted to a universal shape, which we take to be a Zhao profile (Zhao 1996)
\be
\rho\left(r\right)=\rho_s\left(\frac{r}{r_s}\right)^{-\alpha} \frac{1}{\left(1+r/r_s\right)^{\epsilon-\alpha}}. \label{Zhao}
\ee
Here $r_s$ is a characteristic length scale and $\rho_s = \delta_c \ \rho_{crit}$ is a characteristic density, which is equal to a density enhancement $\delta_c$ times the critical density for closure $\rho_{crit}=3H^2/8\pi G$.  The two free parameters $\delta_c$ and $r_s$ can be determined from the halo concentration $c_{\Delta}\left(M_{\Delta}\right)$ and the virial mass $M_{\Delta}$

\be
\delta_c=\frac{\Delta}{3} \ \frac{c_{\Delta}^3}{ln\left(1+c_{\Delta}\right)-c_{\Delta}/\left(1+c_{\Delta}\right)}
\ee
\be
r_s=\frac{r_{\Delta}}{c_{\Delta}}=\frac{9.52\times 10^{-2}}{c_{\Delta}}\left(\frac{M_{\Delta}}{M_{\odot}}\right)^{1/3} h^{-2/3} \Delta^{-1/3} \mathrm{kpc}
\ee
where
\be
\Delta\left(\Omega, \Lambda\right) = 178 \left\{ \begin{array}{ll}                        \Omega^{0.30}, & \mbox{if $\Lambda=0$} \\ 
           \Omega^{0.45},  & \mbox{if $\Omega+\Lambda=1$}  \end{array}  . \right. \label{v2}
\ee
Here $r_{\Delta}$ is the virial radius, inside which the average overdensity is $\Delta$ times the critical density for closure, and $M_{\Delta}$ is the mass within $r_{\Delta}$.  For any particular cosmology, the concentration $c_{\Delta}\left(M_{\Delta}\right)$ is a function of the virial mass which results from the fact that dark halo densities reflect the density of the universe at their formation epoch, and smaller mass halos collapse earlier in hierarchical structure formation.  Hence, for CDM power spectra, $c_{\Delta}\left(M_{\Delta}\right)$ is a decreasing function of $M_{\Delta}$.  ENS (2001) have carried out an extensive suite of N-body simulations to characterize the dependence of $c_{\Delta}\left(M_{\Delta}\right)$ on the cosmological parameters by fitting their simulated dark halos to an NFW profile, which is eqn. (\ref{Zhao}) with $\alpha=1$ and $\epsilon=3$.  We have used their publicly available routine to calculate $c_{\Delta}\left(M_{\Delta}\right)$ for the $\Lambda$CDM cosmology quoted in \S 1.  Strictly speaking, these concentrations apply only to the NFW profile; however, we use them to characterize the more general profile in eqn. (\ref{Zhao}), since $c_{\Delta}\left(M_{\Delta}\right)$ should not change significantly if the inner profile slope $\alpha$ is instead taken as a free parameter.
	
 	Given the characteristic radius and density, $r_s$ and $\rho_s$, the characteristic velocity dispersion $v_s$ can be calculated from $v_s^2 =\mu G\rho_s r_s^2$ where $\mu\equiv\frac{2\pi}{\left(3-\alpha\right)\left(\alpha-1\right)}$.  The density profile in eqn. (\ref{Zhao}), will give a circular velocity profile, $v_{circ}\left(r\right)$, which increases outward from the center passing through a maximum at $r\sim r_s$ and then decreases.  The circular velocity of a dark halo is conventionally taken to be the value at this maximum, since $r_s$ is the only length scale in the system.   For the NFW profile, this maximum occurs at $r\approx 2 r_s$, but for our more general profile the location of the maximum will depend on $\alpha$ and $\epsilon$.  We define the circular velocity of a dark halo $v_{circ}$, to be the maximum of the NFW profile (eqn. (\ref{Zhao}) with $\alpha=1.0$ and $\epsilon=3.0$), again assuming that $v_{circ}$ will not change significantly if $\alpha$ is allowed to vary.
\be
v_{circ}=3.13 \times 10^{-3} \left(\frac{M_{\Delta}}{M_{\odot}}\right)^{1/3}\left(\frac{c_{\Delta}}{ln\left(1+c_{\Delta}\right)-c_{\Delta}/\left(1+c_{\Delta}\right)}\right)^{1/2} h^{-1/3}\Delta^{1/6} \ \mathrm{km} \ \mathrm{s}^{-1}
\ee

\section{DENSITY CUSP ENHANCEMENT BY DISSIPATIVE COLLAPSE OF BARYONS}

	The dissipative infall of baryonic matter will strongly perturb the underlying dark matter distribution, pulling it inward and resulting in a steeper density cusp in the innermost regions of the dark halo.  If the dark matter is collisionless and the baryons condense slowly (compared to the dynamical time in the inner halo), the angular momentum, or equivalently $rM\left(r\right)$, is an adiabatic invariant, allowing one to calculate the final density profile of the dark halo given the initial dark matter profile and the final profile of the baryons (Blumenthal et al. 1986; Flores et al. 1993).  

	For our purposes, we are interested in the density profile of the dark halo for $0\lesssim r\lesssim r_{\ast}$, where $r_{\ast}$ is defined as the radius of optical depth unity: 

\be
r_{\ast}\equiv r_s \tau_{s}^{\frac{1}{\beta-1}},
\ee
with the optical depth given by eqn. (\ref{tau}a) and $\tau_s$ by eqn. (\ref{tau}b).  For the density profiles given by eqn. (\ref{NFW}), the relaxation time goes to zero at the center.  This sets a limit to the minimum radius, $r_H$, for which the system can be considered collisionless, where the relaxation time equals the age of the halo.  For the range of scales and parameters we consider, $r_{\ast} \ll r_H$; hence, dark matter is highly collisional in the region where the density profile is desired, and the adiabatic approximation breaks down: collisions will scatter dark matter particles onto orbits of different angular momentum, breaking the invariance of $rM\left(r\right)$.  For regions of the halo with $\tau\left(r\right) \sim 1$ $(r\sim r_{\ast})$ the adiabatic approximation should hold reasonably well; however, for regions with $\tau\left(r\right) \gg 1$ ($r\ll r_{\ast}$), the dark matter behaves as a fluid, so that the equations of hydrostatics must be employed.  

	In what follows we determine the steepening of the inner density cusp in the collisionless and highly collisional (fluid) limits, and find the slope of the final dark matter density cusp to be identical.  We do not know of a simple approximation to determine the cusp in the region where the dark matter is moderately collisional, but expect the density profile to interpolate smoothly between the inner (fluid) and outer (collisionless) regions of the halo, as the cusp slope will be the same.  In the following derivations we drop all numerical constants and consider only the scaling of various quantities with $r$.  Subscripts $i$ and $f$ are used where appropriate to distinguish the initial and final states.  In the initial state, the dark matter and baryons are assumed to be well mixed, so that the density profile of the baryons will be parallel to that of the dark matter.  In the final state, the baryons are assumed to dominate the total mass (dark matter plus baryons) $M_f\left(r_f\right)$ in the inner region where the dark matter profile is desired, and this final total mass distribution is taken to be $M_f\sim r_f^{3-\xi}$ corresponding to a total density profile $\rho_{f}\sim r_f^{-\xi}$.  We assume the dark matter density cusp is $\rho_i \sim r_i^{-\alpha}$ before condensation and $\rho_f \sim r_f^{-\alpha^{\prime}}$ after, and determine $\alpha^{\prime}=f\left(\alpha,\xi\right)$.

	For the collisionless case, we make the simplifying approximation that the dark matter particles are on circular orbits, which is justified since there is more phase space available for nearly circular orbits than for radial ones (Blumenthal et al. 1986; Flores et al. 1993; Navarro, Frenk \& White 1996).  Conservation of dark matter implies that 
\be
\rho_i r_i^2 dr_i=\rho_f r_f^2 dr_f \ \Rightarrow r_i^{3-\alpha} \sim r_f^{3-\alpha^{\prime}} . \label{masscon}
\ee
Conservation of angular momentum gives
\be   
r_iM_i\left(r_i\right)=r_f M_f\left(r_f\right) \sim r_f^{4-\xi} \Rightarrow r_i^{4-\alpha} \sim r_f^{4-\xi}. 
\ee
Combining these two results we find
\be	
\alpha^{\prime}=\frac{\alpha+3\xi-\alpha\xi}{4-\alpha}.
\ee

	In the fluid limit, we assume the dark matter is in hydrostatic equilibrium in the initial and final states.  We neglect heat transfer, so that the entropy of the dark matter at fixed mass shell is invariant
\be
S_i\left(r_i\right)=S_f\left(r_f\right), \label{enteq}
\ee
where $S$ is given by
\be
S\left(r\right)=\frac{P\left(r\right)}{\rho^{5/3}\left(r\right)} \label{entropy}
\ee
for a monatomic ideal gas, and $r_i$ and $r_f$ are related by mass conservation (eqn. (\ref{masscon}).  These assumptions are valid provided the condensation of baryons takes place on a timescale slow relative to the dynamical time, but fast compared to the heat conduction timescale (eqn. (\ref{ttherm})). If $t_{B}$ is the condensation time, these approximations hold where $t_{dyn}\left(r\right)\lesssim t_{B}\lesssim \tau\left(r\right) t_{dyn}\left(r\right)$.  Hydrostatic equilibrium of the dark matter component implies  
\be
\frac{1}{\rho}\frac{dP}{dr} \sim \frac{M\left(r\right)}{r^2} \Rightarrow P \sim \rho M r^{-1};
\ee
or
\be
P_i \sim r_i^{2-2\alpha} \ ; \ P_f \sim r_f^{2-\xi-\alpha^{\prime}}.
\ee
Combining with eqns. (\ref{enteq}) and (\ref{entropy}) gives
\be
r_i^{2-\alpha/3} \sim r_f^{2-\xi+2\alpha^{\prime}/3},
\ee
and mass conservation (eqn. (\ref{masscon}) then implies
\be
\alpha^{\prime}=\frac{\alpha+3\xi-\alpha\xi}{4-\alpha}, \label{alphaprime}
\ee 
which is identical to the collisionless case.

\begin{figure}
\plotone{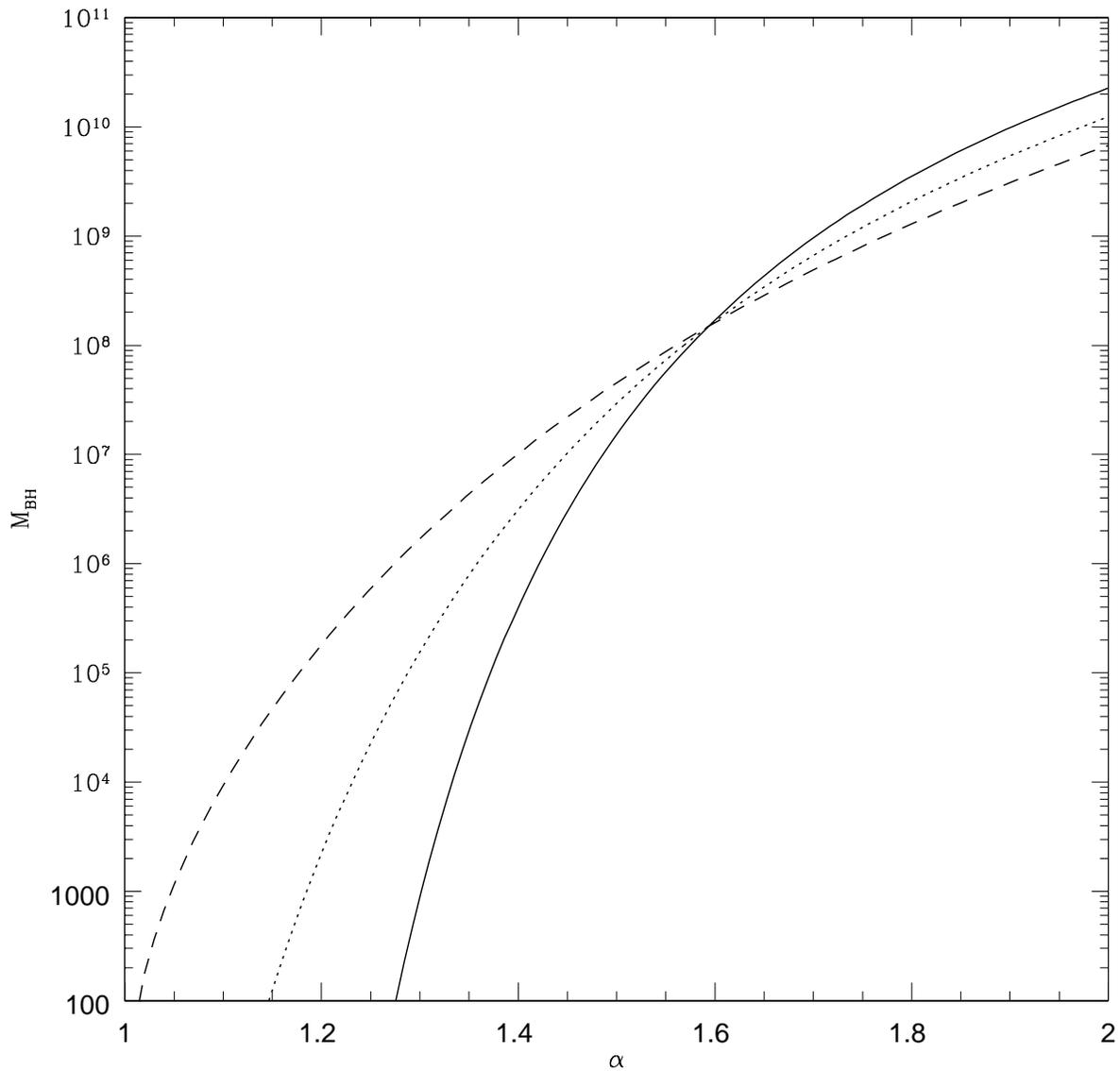}
\caption[Black Hole Mass versus Inner Profile Exponent]{Black hole mass grown from optically thick accretion as a function of the inner density profile exponent $\alpha$ for different values of $a$ (the velocity dependence of the cross section) for a Milky Way sized dark halo. Solid, dotted, and short-dashed curves are for $a=0, 0.5, 1$ respectively. The dark matter interaction strength has been set to $\sigma_1v_{100}^a=1$. The $M_{BH}$ is extremely sensitive to $\alpha$ for $a=0-1$.}
\label{bhalpha}
\end{figure}

	For the most probable value of $\alpha=1.3$, and taking $\xi=2$ for a final flat rotation curve, typical of normal bulge dominated galaxies, $\alpha^{\prime}=1.74$.  Note that in deriving our black hole constraint from bulge systems in \S 5 we use the initial state values of $r_s$, $\rho_s$, and $v_s$, since compression will only change these quantities by factors of order unity, resulting in an insignificant change in $M_{BH}$ compared to the change of several orders of magnitude caused by the steeper cusp $\alpha^{\prime}$ (see Figure \ref{bhalpha}).

{}

\newpage


\begin{table*}
\begin{center}

TABLE 1

Formation Redshifts and Times for Dark Halos

\

\begin{tabular}{ccc}
\tableline\tableline
$M_{halo}$ & $z_f$ & $t_f/t_{H}$ \\
($M_{\odot}$) & &\\
\tableline

$10^8$  & 1.83 & 0.26 \\
$10^9$  & 1.61 & 0.29 \\
$10^{10}$ & 1.38 & 0.33 \\
$10^{11}$ & 1.15 & 0.39 \\
$10^{12}$ & 0.93 & 0.45 \\
$10^{13}$ & 0.72 & 0.52 \\
$10^{14}$ & 0.53 & 0.61 \\
$10^{15}$ & 0.36 & 0.71 \\
\tableline
\end{tabular}
\end{center}
NOTE.$-$ Formation times and redshifts for dark halos from the dwarf galaxy to cluster scale in a $\Lambda$CDM cosmology.  These are calculated from the Lacey \& Cole (1993) formation time distribution.  $M_{halo}$ is the virial mass of the dark halo, $z_f$ the formation redshift, and $t_f/t_H$ the ratio of the formation time to the age of the universe.

\end{table*}

\newpage
 
\newpage

\newpage

\newpage

\newpage

\newpage

\end{document}